\documentclass[preprint,5p]{elsarticle}

\pdfoutput=1

\usepackage{color}

\usepackage{cite}

\usepackage{amsmath}
\usepackage{amssymb}
\usepackage{specfun}
\usepackage{phys}     

\usepackage{mystyle}

\graphicspath{{figures/}}

\newcommand{\spanvs}{\mathrm{span}}
\newcommand{\im}{\mathrm{im}}

\begin{document}

\begin{frontmatter}


\title{Study of the non-linear eddy-current response in a ferromagnetic plate: theoretical
analysis for the 2D case}

\author[CEA]{Anastassios~Skarlatos}
\author[UoWM]{Theodoros~Theodoulidis}
\address[CEA]{A. Skarlatos is with the CEA, LIST, F-91191 Gif-sur-Yvette cedex, France.}
\address[UoWM]{T. Theodoulidis is with the Department of Mechanical Engineering, University of 
Western Macedonia, Bakola \& Sialvera, 50100 Kozani, Greece.}	




\begin{abstract}
The non-linear induction problem in an infinite ferromagnetic pate is studied theoretically by 
means of the truncated region eigenfunction expansion (TREE) for the 2D case. The non-linear 
formulation is linearised using a fixed-point iterative scheme, and the solution of the 
resulting linear problem is constructed in the Fourier domain following the TREE formalism. 
The calculation is carried out for the steady-state response under harmonic excitation and
the harmonic distortion is derived from the obtained spectrum. This article is meant to be the
theoretical part of a study, which will be complemented by the corresponding experimental work 
in a future communication. 
\end{abstract}


\end{frontmatter}


\section{Introduction}

In material characterization applications, the tested specimen is subject to a strong 
electromagnetic field in order to trigger its non-linear behaviour. The main  interest in 
these techniques relies on the fact that the magnetic properties of ferromagnetic media, 
especially their hysteretic characteristics, are strongly related to their micro-structure, 
and hence they provide an indirect link for the assessment of properties like mechanical 
strength, presence of residual stresses, etc., which are usually accessible through 
destructive tests or other complicated and expensive techniques. 

In case of planar specimens like strips or plates, the two main methods for establishing the 
excitation field are using air-cored coils, located above or at both sides of the piece, or 
via a closed magnetic circuit (yoke), which is brought in contact with the specimen interface.

The simulation of the inspection procedure with either methods requires the solution of a 
non-linear, hysteretic problem. The standard way of treating this problem is via successive 
linearisation using either the fixed-point method (also known as polarization or Picard-Banach
method) or the Newton-Raphson scheme and the application of a numerical technique for the 
solution of the resulting linear problem at each iteration. Considerable progress has been 
made the past years in the development of such solvers based on the finite elements method 
(FEM) \citep{albanese_transmag92, biro_transmag06}, the finite integration technique (FIT) 
\citep{drobny_transmag00, clemens_compel11, dutine_intjnumermodel17} or the integral 
equation approach \citep{albanese_transmag96, circ_transmag07, daquino_transmag14}. The 
relevant literature is vast, and the above list should be understood only as indicative.

The main inconvenience in using these techniques is that they rely on the application of a
volume mesh, with a large number of degrees of freedom (dofs), which results in the repeated 
inversion of a large (sparse or full, depending on the formulation) system of linear equations.
To overcome this drawback, sophisticated techniques using semi-explicit schemes for the 
minimization of linear system inversions have been proposed \citep{clemens_compel11, 
dutine_intjnumermodel17}. Another approach to cope with the raised computational effort is to
resort to hardware acceleration e.g. using parallel and/or GPU adapted implementations 
\citep{daquino_transmag14, boerm_comput06, rubinacci_jcompphys09}. Note here, that especially 
in the case of the integral equation approach, whose main drawback is related to the 
limitations in terms of CPU time and memory due to the full matrices involved, the new 
generation of solvers based on the sparsification techniques with parallel and/or GPU 
adapted implementations, as the ones mentioned in \citep{boerm_comput06, 
rubinacci_jcompphys09}, has allowed the efficient treatment of large, previously intractable
problems.

An additional difficulty is 
linked with the existence of steep field gradients inside the ferromagnetic materials, which 
raise increased demands in terms of grid resolution, thus reducing the robustness of the 
solution and making human expertise indispensable in order to assure the validity of the 
results. 

The above mentioned drawbacks can be partly avoided if we are willing to sacrifice the 
versatility of generic numerical solvers for the favour of more case-dependant modal 
approaches. Indeed, since the majority of eddy-current inspection/evaluation configurations 
involve relatively simple geometries that are amenable to semi-analytical solutions, this 
approach can be a valuable aid to the analysis, owing to the very convenient computational 
times and the absence of a computational mesh\footnote{In the case of the non-linear solver,
the spatial discretisation of the magnetisation cannot be avoided, yet the mesh used for its 
evaluation is restricted inside the ferromagnetic material, and it impacts the solution only 
indirectly as it will become clear by the analysis.}.

In this article, the non-linear eddy-current response of an infinite ferromagnetic plate to 
two coaxial air-cored cylindrical coils located at its both sides will be studied by means of 
a modal approach. The considered configuration presents an important practical interest since 
it stands for one of the two most important experimental set-ups used in material evaluation 
applications. The present contribution copes with the theoretical analysis of the problem. An 
experimental study and comparison with the theoretical results presented here will be the 
subject of a work in prepare.

The developed solution follows the approach of the polarisation technique, that is, the problem
is linearised assuming a constant permeability in the ferromagnetic piece, and the unknown 
magnetic polarisation is determined via fixed-point iterations. The thus resulting linear 
problem is a multilayer eddy-current problem with a distributed magnetic source exceeding the 
domain of the ferromagnetic piece, which is treated by means of the truncated region 
eigenfunctions expansion (TREE) \citep{theobook, skarlatos_transmag15}. A simpler version of
the proposed approach has been applied for the study of the 1D problem 
\citep{skarlatos_transmag16}. 

The paper is organized as follows. First the general scheme for the linearisation of the state
equation is presented without any specific reference to the solution method. The treatment of 
the resulting linear multilayer problem follows in section III. A discussion on the numerical 
issues arising from both the modal solution and the application of the non-linear operator
is presented in section IV. The results of the proposed method are then compared with the ones 
obtained via a FIT implementation in time domain based on implicit Euler scheme.

Only the case of harmonic excitation is considered in this work. The article follows the idea
of developing the solution in a series of the first harmonics and treating each harmonic
separately in the frequency domain \citep{biro_transmag06, circ_transmag07, daquino_transmag14}.
The more general case of the transient response calculation for finite duration (finite energy) 
signals needs a different treatment, and will be studied in a future work.

\section{The non-linear formulation}

\subsection{Problem statement}

\begin{figure}[h]
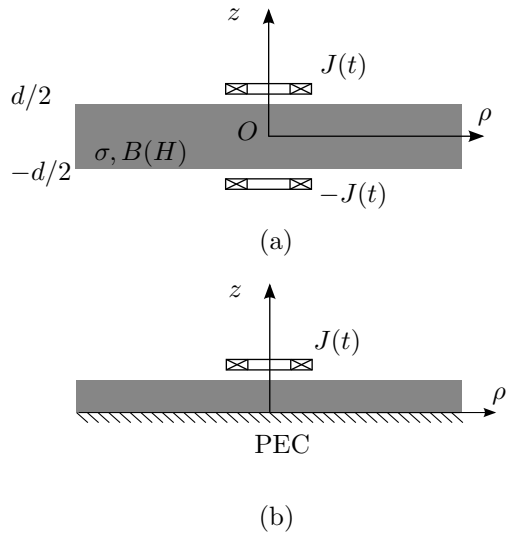
\centering
\caption{Ferromagnetic plate sandwiched between two coils fed with currents of opposite sign: 
(a) original configuration and (b) equivalent problem.}
\label{fig:PlateConf}
\end{figure}

Let us consider an infinite ferromagnetic plate of thickness $d$ excited by a pair of coaxial 
coils whose axis is normal to the plate as shown in \figref{fig:PlateConf}. The two coils are 
located at the two opposite sides of the plate, and they are fed with opposite currents of the 
same amplitude. This specific excitation mode creates a strong, nearly stationary, tangential 
magnetic field inside the plate, thus maximizing the magnetization effects inside the 
material. Using the common practice when dealing with symmetrical configurations, only 
the upper half of the geometry needs to be considered, the anti-symmetry of the induced 
eddy-current flow being imposed via a perfectly conducting electric boundary condition (PEC) 
passing from the middle of the plate as shown in \figref{fig:PlateConf}b. The solution to a 
problem with the given symmetry is referred to in the literature as odd-parity solution 
\citep{skarlatos_transmag15}. In the rest of the text, we shall refer to the plate volume as 
region 1, whereas the air domain above the plane will be named as region 2.
%

The plate has non-zero conductivity $\cond$, and its magnetic properties are described via the
constitutive equation 
%
\begin{equation}
\Bflux = \permb_0[\Hfield + \Magn(\Hfield)]
\label{eq:ConsRelVec}
\end{equation}	
where $\permb_0$ is the magnetic permeability of the free space, and $\Magn$ stands for the 
magnetization of the material, which comprises the non-linear effects. As analysed in 
\citep{skarlatos_transmag16}, it is often beneficial from the computational point of view to
introduce an effective relative permeability $\permb_r$ greater than one. 
\Eqref{eq:ConsRelVec} then can be rearranged as follows
%
\begin{equation}
\Bflux = \permb_r\permb_0\Hfield + 
\permb_0\left[\Magn(\Hfield) + \left(1-\permb_r\right)\Hfield\right].
\label{eq:ConsRelIEf}
\end{equation}
As it becomes clear from \eqref{eq:ConsRelIEf}, the role of the effective permeability is to 
adjust the slope of the linear part of the constitutive equation in order to approach the 
actual slope of the non-linear curve, and hence to improve convergence. The exact value of 
$\permb_r$ is however subject to the constrain  $\permb<2\permb_{min}$, $\permb_{min}$ being 
the minimum of the differential permeability, to assure the stability of the iterative scheme 
\citep{hantila_revroum75}. Practically, the choice of $\permb$ is a compromise between 
convergence speed and stability of the algorithm. For the rest of the article, the convergence 
will be considered as guaranteed without further investigation.

We define $\Ipol\fun{\Hfield} := \permb_0\left[\Magn\fun{\Hfield} + 
\left(1-\permb_r\right)\Hfield\right]$, which will be referred to as effective magnetic 
polarization, or simply magnetic polarization, for the rest of the text. For $\permb_r=1$ it 
reduces to the usual definition of the magnetic polarization, which is the magnetization of 
the material scaled by the magnetic permeability of the free space\footnote{Both variables 
describe exactly the same physical quantity; the reason for the introduction of the magnetic 
polarization is mere mathematical convenience. Notice that in the cgs system of units 
$\permb_0=1$ and hence the magnetic polarization and the magnetization coincide.}. 

\subsection{The governing equation}

Problems with rotational symmetry can be scalarised by introducing the magnetic vector 
potential, defined as
%
\begin{equation}
\Bflux = \nabla\times\Apot
\label{eq:ApotDef}
\end{equation}
and making the following ansatz
%
\begin{equation}
\Apot = A\uvect{\phi}
\label{eq:ApotAnsatz}
\end{equation}
where $\uvect{\phi}$ is the unit vector along the azimuthal direction.

Upon substitution of \eqspairref{eq:ApotDef}{eq:ApotAnsatz} into the Maxwell's equations and 
taking the constitutive relation \eqref{eq:ConsRelIEf} into account, we obtain the diffusion 
equation for the magnetic potential
%
\begin{equation}
\nabla^2 A - \permb\cond\frac{d A}{dt} = \permb_0 J - \uvect{\phi}\cdot\nabla\times\Ipol.
\label{eq:InhomAHelmAir}
\end{equation}
with $\permb = \permb_r\permb_0$, $J$ being the coil current (which by hypothesis has only 
azimuthal component) and $\Ipol$ the magnetic polarization defined above.

To simplify the notation, we define the linear differential operator 
$\mathcal{D} := \nabla^2 - \permb\cond d_t$, and the non-linear material operator 
$\mathcal{I} := \uvect{\phi}\cdot\nabla\times\Ipol\bfun{\bullet\,\uvect{\phi}}$, which both act 
on the scalarised potential $A$. The solution of \eqref{eq:InhomAHelmAir} can be written
formally as
%
\begin{equation}
A = 
\left[
\vect{1} - \mathcal{D}^{-1}\mathcal{I}
\right]^{-1} \mathcal{D}^{-1} \permb_0 J.
\label{eq:FormalSol}
\end{equation}
where $\vect{1}$ stands for the unit operator.

The bracketed term is a non-linear operator and as such it cannot be inverted analytically. To
evaluate \eqref{eq:FormalSol} approximately, we develop the bracketed expression in Taylor 
series
%
\begin{equation}
A = 
\left[
\vect{1} - \mathcal{D}^{-1}\mathcal{I} +
\left(\mathcal{D}^{-1}\mathcal{I}\right)^2 + \ldots
\right] \mathcal{D}^{-1}\permb_0 J.
\label{eq:FormalSolSeries}
\end{equation}
The expression in \eqref{eq:FormalSolSeries} is an alternative way of writing the fixed-point
iterative scheme. More precisely, the total solution is obtained by successive application of 
the non-linear operator $\mathcal{D}^{-1}\mathcal{I}$ to the coil response inside the 
linearised medium, namely
%
\begin{equation}
A_{i+1} = A_{0} + \mathcal{D}^{-1}\mathcal{I} A_{i}
\label{eq:FixedPoint}
\end{equation} 
for $i=0,1,2,\ldots$ and $A_{0} = \permb_0\mathcal{D}^{-1}J$. 

An important feature of the fixed-point iterative scheme that should be emphasised here is 
that in contrast from other approaches, like Newton, the uniqueness of the solution, at least 
in the ideal case, can be proved, under reasonable hypotheses on the behaviour of the magnetic 
constitutive relationship. The approximate solution, in this case, does not suffer of the 
local minima that cannot be excluded in other iterative approaches.

In the previous discussion, the general scheme for tackling the non-linear problem has been
presented without any specific reference of how to solve $\permb_0\mathcal{D}^{-1}J$ and 
evaluate the expression \eqref{eq:FixedPoint}. Indeed, both tasks can be carried out using any 
established numerical scheme. In the the following paragraphs, we will present a semi-analytical 
(modal) approach for treating both tasks. The main advantage of this approach is the 
diagonalisation of the linear operator $\mathcal{D}$, which allows fast computation of the 
fixed-point iterations.

\section{Modal solution of the linearised problem}

Only the case of harmonic excitation will be considered in this work, i.e. the time dependency 
of the coil current is taken equal to $J\fun{t} = J_0 \sin\fun{i\omega_0 t}$, $\omega_0$ being 
the angular frequency. The frequency is assumed sufficient low in order the quasi-static 
approximation to be valid.

Due to the non-linearity of the material, the final solution will contain, beside the basic 
harmonic $\omega_0$, all its odd multiples $\omega_p = p\omega_0$, with $p=1,3,\ldots,\infty$. 
All the time-dependant physical quantities of the problem are thus given by the sum
%
\begin{equation}
f\fun{\rpos,t} = \sum\limits_{p=-\infty}^{\infty} f_p\fun{\rpos} e^{i\omega_p t}, \;
p=\pm 1,\pm 3,\ldots,\infty
\end{equation}
where the development coefficients are given by the integrals
%
\begin{equation}
f_p\fun{\rpos} = \frac{1}{T}\int\limits_{-\infty}^{\infty} f\fun{\rpos,t} e^{-i\omega_p t} dt
\end{equation}
$T$ being the time period of the signal (equals the period of the excitation signal 
$\omega_0/2\pi$). In practice, the spectrum $f_p$ decreases very rapidly with the frequency, 
and hence only a very small number of harmonics is sufficient to provide an excellent 
accuracy.

In order to avoid unnecessary notation complexity, the harmonic index $p$ will  be dropped 
henceforth. All following relations are understood as being applied for the $p$th harmonic
unless otherwise specified.

\subsection{Calculation of the coil response: $\permb_0\mathcal{D}^{-1}J$}

We seek solution to the linear, odd-parity, eddy-current problem of a cylindrical coil over an 
infinite plate excited by an harmonic current. This is a typical eddy-current testing situation,
whose solution is very well studied in the literature \citep{theobook,skarlatos_transmag15}. 
This section will thus serve as a reminder of the basic results, which will be presented 
without proof.

Following the common practice of the TREE approach, the computational domain is truncated at 
$\rho=\rho_L$ taking advantage of the diffusive character of the solution. Assuming that the 
$\rho_L$ is sufficiently large in order the electromagnetic field to be negligible, we are 
free to choose the type of condition at the truncation boundary that is more convenient for 
the analysis. Let us consider a perfect electric conducting (PEC) condition at the truncation
boundary.
 
The solution for the magnetic potential reads
%
\begin{equation}
A^{(1)}\fun{\rho,z} = 
\sum\limits_{\ell=1}^\infty
\besselj{1}\fun{\kappa_\ell \rho}
\left[
C_\ell^{(s)} e^{\kappa_\ell z} + D_\ell^{(d)} e^{-\kappa_\ell z}
\right]
\label{eq:A1coil}
\end{equation}
in the air region above the plate, and
%
\begin{equation}
A^{(2)}\fun{\rho,z} = 
\sum\limits_{\ell=1}^\infty
\besselj{1}\fun{\kappa_\ell \rho} 
C_\ell^{(c)} \sinh\fun{v_\ell z}
\label{eq:A2coil}
\end{equation}
inside the plate, with the eigenvalues $\kappa_\ell$ being determined by the boundary 
condition at the truncation surface
%
\begin{equation}
\besselj{1}\fun{\kappa_\ell \rho_L} = 0
\label{eq:kappaDef}
\end{equation}
and $v_\ell$ satisfying the dispersion relation
\begin{equation}
v_\ell^2 = \kappa_\ell^2 + k_p^2.
\label{eq:vDef}
\end{equation}
$k_p^2$ stands for the dispersion coefficient in the plate and for the $p$th harmonic: 
$k_p^2 = i\omega_p\mu\sigma$. Notice that in air $k_p^2=0$, which yields $v_\ell = \kappa_\ell$.

$C_\ell^{(s)}$ are the development coefficients for the solution of the coil in the absence of
the plate, and they depend only upon the coil geometry and the excitation current. Their 
expression for the air region between the coil and the plate are given by\footnote{Similar 
expressions can be obtained for the remaining parts of the space. Yet, they will not be needed 
in the context of the present analysis, and hence they will not be considered here. The 
interested reader is referred to the literature for the full solution\citep{theobook}.}
%
\begin{align}
C_\ell^{(s)} &= -4\permb_0 \iota_0  
\sinh\fun{\frac{\kappa_n l}{2}}
\frac{\chi\fun{\kappa_n\rho_{in},\kappa_n\rho_{out}}}
{\kappa_n^5 \left[\rho_L\besselj{1}\fun{\kappa_n \rho_L}\right]^2}
\nonumber\\
&\times
\besselj{m}\fun{\kappa_n \rho_0} e^{-\kappa_n z_0}
\end{align}
where $\rho_{in,out}$ stand for the coil inner and outer radius, respectively, $l$ is the coil
thickness, and $\iota_0$ is the current density across its cross-section. The $\chi$ function 
stems from the integration over the coil cross-section and it has a closed-form expression in 
terms of Struve functions $\struve{n}\fun{x}$: $\chi\fun{x_1,x_2} = \frac{\pi}{2} 
[x\besselj{0}\fun{x}\struve{1}\fun{x} - x\besselj{1}\fun{x}\struve{0}\fun{x}]_{x_1}^{x_2}$.

The expressions for the reflection and transmission coefficients $D_\ell^{(d)}$ and 
$C_\ell^{(c)}$ read
%
\begin{align}
D_\ell^{(d)} = 
\frac{2\permb_r\kappa_\ell}{\permb_r\kappa_\ell\sinh\fun{v_\ell d/2} + v_\ell\cosh\fun{v_\ell d/2}}
e^{\kappa_\ell d/2} C_\ell^{(s)}
\end{align}
and
%
\begin{align}
C_\ell^{(c)} = 
\frac{\permb_r\kappa_\ell\sinh\fun{v_\ell d/2} - v_\ell\cosh\fun{v_\ell d/2}}
{\permb_r\kappa_\ell\sinh\fun{v_\ell d/2} + v_\ell\cosh\fun{v_\ell d/2}}
e^{\kappa_\ell d/2} C_\ell^{(s)}
\end{align}
respectively.

\subsection{Solution of the update equation: $\mathcal{D}^{-1}\mathcal{I}A_{i}$}

Our task here is to calculate the potential solution at the $(i+1)$th iteration, assuming that 
the magnetic polarization is given (its value is obtained by applying the material operator to
the solution of the previous iteration). Formally speaking, we deal with the solution of the
inhomogeneous Helmholtz equation, where the support of the right hand side (excitation) term 
is restricted to the domain of the plate:
%
\begin{equation}
\left(\nabla^2  - k_p^2\right) \Delta A_{i+1} = -\uvect{\phi}\cdot\nabla\times\Ipol_i
\label{eq:UpdateEq}
\end{equation}
where $\Delta A_{i+1} = A_{i+1} - A_{0}$. The solution of \eqref{eq:UpdateEq} in the interior 
of the plate can be expanded in two, mutually orthogonal, vector spaces
\begin{equation}
\Delta A^{(2)} = \sum\limits_{\ell,n} c_{\ell n} u_{\ell n} + 
\sum\limits_{\ell} C_\ell^{(I)} U_{\ell}
\label{eq:UpdateEqDev}
\end{equation}
which span the image and the kernel of the differential operator $\mathcal{D}$ respectively,
i.e. it is $\spanvs\{u_{\ell n}\} = \im\bfun{\mathcal{D}}$ and $\spanvs\{U_n\} = 
\ker\bfun{\mathcal{D}}$, with $\im\bfun{\mathcal{D}} \perp \ker\bfun{\mathcal{D}}$ (the double
index for the image subspace stems from the two dimensions of the image space). Differently 
stated, $u_{\ell n}$ yields a special solution for the non-homogeneous equation, whereas $U_n$
assures the uniqueness of the total solution according to Fredholm's alternative theorem. 
Indeed, $u_{\ell n}$ will introduce a discontinuity at the plate interface, which will be 
revealed by $U_n$ as it will be shown below.

A convenient choice for $u_{\ell n}$ and $U_n$ bases for the expansion of the solution is the 
eigenfunction basis of the Helmholtz operator, which for the given symmetry reads
%
\begin{equation}
u_{\ell n}\fun{\rho,z} = 
\besselj{1}\fun{\kappa_\ell \rho}
\left\{
\begin{array}{ll}
z, \hspace{1ex}  & n=0 \\ 
\sin\fun{\alpha_n z},\;  & n=1,2,\ldots,\infty
\end{array}
\right.
\label{eq:ImBasis}
\end{equation}
for the image subspace, and
%
\begin{equation}
U_{\ell}\fun{\rho,z} = 
\besselj{1}\fun{\kappa_\ell \rho}
\sinh\fun{v_\ell z}
\label{eq:KerBasis}
\end{equation}
for the kernel subspace, where $\kappa_\ell$ and $v_\ell$ are the same eigenvalues with the 
ones considered for the calculation of the coil response (since we deal with the same boundary 
conditions), and the are thus given by \eqref{eq:kappaDef} and \eqref{eq:vDef}, respectivelly. 
The values of $\alpha_n$ depend upon the condition at the plate interface $z=d/2$, which is 
taken to be of PEC-type (again an arbitrary choice). Together with the zero order term $z$,
they establish the basis completeness. The PEC condition yields for $\alpha_n$
%
\begin{equation}
\alpha_n = 2n\pi/d
\end{equation}
with $n=1,2,\ldots,\infty$.

We need now to determine the development coefficients $c_{\ell n}$ and $C_\ell^{(I)}$. 
Substitution of \eqref{eq:UpdateEqDev} upon the Helmholtz equation \eqref{eq:UpdateEq} yields
%
\begin{equation}
\sum\limits_{\ell=1}^{\infty}\sum\limits_{n=0}^{\infty}
\left(
\kappa_\ell^2 + \alpha_n^2 + k_p^2 
\right) 
c_{\ell n} u_{\ell n}\fun{\rho,z} = 
\uvect{\phi}\cdot\nabla\times\Ipol
\label{eq:SpecSol}
\end{equation}
with $\alpha_0 = 0$. Observing \eqref{eq:SpecSol} at $z=d/2$, and taking into account the 
orthogonality of the basis, we obtain the first relation for the zero-order $c_{\ell 0}$ 
coefficients
%
\begin{equation}
\left(
\kappa_\ell^2 + k_p^2 
\right) 
c_{\ell 0} = 
\frac{2}{E^2}
\inprod{u_{\ell 0}\fun{\rho,d/2}, \uvect{\phi}\cdot\nabla\times\Ipol}_{z=d/2} 
\label{eq:c0Coeffs}
\end{equation}
where $\inprod{\bullet,\bullet}_{z=d/2}$ denotes the inner product for the radial part of 
$u_{\ell 0}$, namely
%
\begin{equation}
\inprod{f,g}_{z=d/2} := \int\limits_{0}^{\rho_L} \rho f\fun{\rho} g\fun{\rho} d\rho
\end{equation}
and $E=\rho_L\besselj{0}\fun{\kappa_\ell\rho_L}$ is the normalization coefficient of the
Bessel function basis.

The remaining coefficients ($n\ne 0$) are obtained by considering the volume of the plate,
i.e.
%
\begin{align}
\left(
\kappa_\ell^2 + \alpha_n^2 + k_p^2 
\right) 
c_{\ell n} 
&= 
\frac{4}{E^2 d}
\inprod{u_{\ell n}\fun{\rho,z} ,\uvect{\phi}\cdot\nabla\times\Ipol} 
\nonumber\\
&-
\frac{4}{E^2 d}
c_{\ell 0}\inprod{u_{\ell n}\fun{\rho,z} ,z}
\label{eq:cCoeffs}
\end{align}
with the inner product $\inprod{\bullet,\bullet}$ defined as
%
\begin{equation}
\inprod{f,g} := 
\int\limits_{0}^{\rho_L}\int\limits_{0}^{d/2} \rho f\fun{\rho,z} g\fun{\rho,z} d\rho dz.
\end{equation}

Both \eqref{eq:c0Coeffs} and \eqref{eq:cCoeffs} involve spatial derivatives of the magnetic 
polarization, which is obtained after application of the non-linear material operator to the
magnetic field solution of the previous step. Practically, $\Ipol$ is evaluated at a finite 
number of grid points, and the integrals of the \eqspairref{eq:c0Coeffs}{eq:cCoeffs} must then 
be computed numerically. It is thus more convenient from the computational point of view to 
pass the curl operator on the left side of the inner products and perform the derivations 
analytically. This is possible thanks to the hermiticity of the curl operator. It can be shown 
after some standard manipulations that \eqspairref{eq:c0Coeffs}{eq:cCoeffs} reduce to
%
\begin{align}
c_{\ell 0} &=
\frac{2}{\lambda_{\ell 0}^2 E^2}
\int\limits_0^{\rho_L}
\rho
\besselj{1}\fun{\kappa_\ell \rho} \partial_z \left[I_\rho\fun{\rho,z}\right]_{z=d/2} d\rho
\nonumber\\
&+\frac{2}{\lambda_{\ell 0}^2 E^2}
\int\limits_0^{\rho_L}
\kappa_\ell 
\besselj{0}\fun{\kappa_\ell \rho} I_z\fun{\rho,d/2} d\rho
\label{eq:c0Coeffs_red}
\end{align}
for the $n=0$ coefficients, and
%
\begin{align}
c_{\ell n} &= 
-\frac{8\alpha_n}{\lambda_{\ell n}^2 E^2 d}
\int\limits_0^{\rho_L}
\rho\besselj{1}\fun{\kappa_\ell \rho}d\rho
\int\limits_0^{d/2}
\cos\fun{\alpha_n z} I_\rho\fun{\rho,z} dz
\nonumber\\
&+\frac{8\kappa_\ell}{\lambda_{\ell n}^2 E^2 d}
\int\limits_0^{\rho_L}
\rho\besselj{0}\fun{\kappa_\ell \rho}d\rho
\int\limits_0^{d/2}
\sin\fun{\alpha_n z} I_z\fun{\rho,z} dz
\nonumber\\
&+ 
\frac{4\lambda_{\ell 0}^2}{\lambda_{\ell n}^2 d^2 \alpha_n}
\left[
1-(-1)^n
\right]
c_{\ell 0}
\label{eq:cCoeffs_red}
\end{align}
for $n\ne 0$. In the above relations we have set $\lambda_{\ell n}^2 = \kappa_\ell^2 + 
\alpha_n^2 + k_p^2$. $I_\rho$ and $I_z$ stand for the radial and axial component of the 
magnetic polarization respectively. The derivative $\partial_z I_\rho$ in 
\eqref{eq:c0Coeffs_red} cannot be reduced any further and has to be evaluated numerically.

To calculate the homogeneous solution and finalize the construction of the total solution, we 
also need to express the potential in the air region, due to the magnetic polarization of the 
plate. The general expression will be identical with the one for the coil response, which for
the sake of completeness we rewrite here
%
\begin{equation}
\Delta A^{(1)}\fun{\rho,z} = 
\sum\limits_{\ell=1}^\infty
D_\ell^{(I)} \besselj{1}\fun{\kappa_\ell \rho} 
e^{-v_\ell z}.
\label{eq:A1update}
\end{equation}

Application of the continuity relations for $H_\rho$ and $B_z$ at the plate interface leads to
the following linear system of equations
%
\begin{align}
&-\permb_r\kappa_\ell D_\ell^{(I)} e^{-\kappa_\ell d/2} 
= 
v_\ell C_\ell^{(I)}\cosh\fun{v_\ell d/2}  
\nonumber\\
&+ c_{\ell 0} +
\sum\limits_{n=1}^{\infty} 
(-1)^n \alpha_n c_{\ell n}
+ \frac{2}{E^2}
\inprod{\besselj{1}\fun{\kappa_\ell\rho}, I_\rho\fun{\rho,z}}_{z=d/2}
\label{eq:HrCnt}
\end{align}
and
\begin{equation}
D_\ell^{(I)}e^{-\kappa_\ell d/2} = C_\ell^{(I)}\sinh\fun{v_\ell d/2} + c_{\ell 0} d/2.
\label{eq:BzCnt}
\end{equation}
Eliminating $D_\ell^{(I)}$, we obtain the explicit expressions for $C_\ell^{(I)}$
\begin{align}
&\left[
\permb_r\kappa_\ell \sinh\fun{v_\ell d/2} + v_\ell \cosh\fun{v_\ell d/2}  
\right]
C_\ell^{(I)} = 
\nonumber\\ 
&c_{\ell 0} \left(1 + \permb_r\kappa_\ell d/2 \right) + 
\sum\limits_{n=1}^{\infty}(-1)^n \alpha_n c_{\ell n}
\nonumber\\
&+
\frac{2}{E^2}
\inprod{\besselj{1}\fun{\kappa_\ell\rho}, I_\rho\fun{\rho,z}}_{z=d/2}
\end{align}
whereas the corresponding value for $D_\ell^{(I)}$ is obtained directly by \eqref{eq:BzCnt}.

\section{Numerical issues}

In the above analysis all the sums comprise infinite number of terms. In reality, the 
numerical evaluation of the solution requires the truncation of both the Fourier spectrum and
the modal sums in \eqspairref{eq:A1coil}{eq:A2coil} and 
\eqspairref{eq:UpdateEqDev}{eq:A1update}. This is a common issue in modal techniques, and is
exactly the point where the approximation of the method is introduced (it can be seen as the
counterpart of the discussion about the mesh in numerical techniques). There is no generally
applicable rule (exactly just as there is no general rule of how fixing the mesh). A usually
adequate number of modes is of the order of 100-150 per direction. For a more detailed 
discussion the reader is referred to previous works \citep{theodoulidis_transmag10, 
skarlatos_procrsoca12}. The frequency spectrum on the other hand is very rapidly decreasing, 
which means that the number of harmonics that have to be taken into account is of the order 
of ten. Notice that due to the point-symmetry of the $B(H)$ curve, only the odd harmonics
contribute to the spectrum, a very-well known experimental fact.

As already mentioned in the introduction, the claim that the modal approach is mesh-less is 
not entirely true in the case of the non-linear problem. In fact, the application of the
non-linear operator $\mathcal{I}$ to the field solution has to be carried out in spatial and
time domain. This means that the modal solution has to be evaluated numerically at each 
iteration at a number of discrete points and at specific time samples. However, here we do not 
deal with a discretisation with the classical sense, since the solution has already been 
calculated, and hence the applied evaluation mesh has only an indirect impact upon the 
results. The field has just to be sampled as densely as necessary to adequately describe the 
spatial and temporal gradients, in the sense of a Nyquist-like criterion. In the context of 
this work, a uniform orthogonal grid with 1000 points along the radial direction and 100 points 
along the plate thickness has been used. The temporal discretisation was realized using 800
samples per period. More sophisticated sampling schemes like non-uniform grids or radial basis
functions may be considered, yet this is not in the scope of this work.

\section{Results}

The presented formulation has been applied for the solution of the eddy-current evaluation 
problem depicted in \figref{fig:PlateConf}, with a thin strip of soft-steel taken as specimen.
The model results are compared with a reference solution produced using a two-dimensional 
numerical code based on the FIT method. The numerical solution is calculated directly in the 
time domain by means of an implicit time-stepping Euler scheme \citep{clemens_transmag03}. 

\begin{figure}[h]
\centering
\includegraphics[width=7cm]{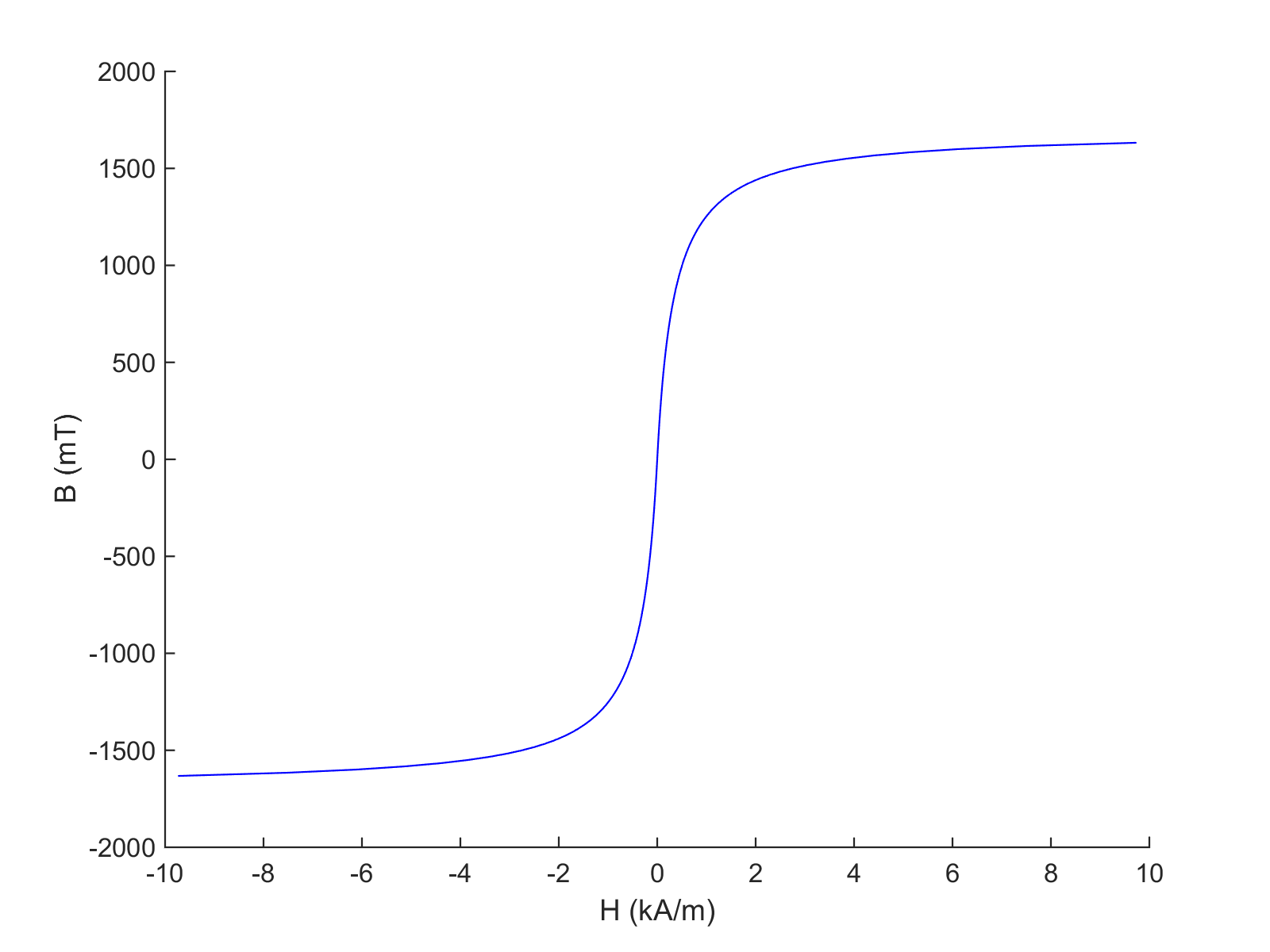}
\caption{Approximated B-H curve of the 1010 steel.}
\label{fig:BHcurve}
\end{figure}

A convenient steel grade for our validation purposes is the 1010 steel, whose behaviour is
with good approximation non-hysteretic. Since it is easier to work with a parametric model
instead of the real experimental curve, the Fr\"ohlich-Kennelly model has been used to 
approximate the material $B(H)$ constitutive relation. It should be noted here that the choice 
of a particular approximation curve does not affect the validation itself, as long as we are 
not comparing the theoretical results with measurements. The only plausible constrain is that 
the considered parametric model must be "non-linear" enough and present the qualitative 
characteristics of a real magnetization curve (i.e. steep slope for low fields, saturation for
field intensities of the order $\sim$ 1-10 kA/m) in order to demonstrate that the model 
reproduces correctly the reference results under these conditions. The explicit expression for 
the Fr\"ohlich-Kennelly model is given by
%
\begin{equation}
B = \frac{H}{\alpha + \beta |H|}.
\label{eq:FroehlichKennely}
\end{equation} 
The $\alpha$ and $\beta$ parameters are usually chosen in order to best fit the experimental 
data. In the present example, the Fr\"ohlich-Kennelly model has been fitted using published
data for the 1010 steel yielding $\alpha=206.42$ and $\beta=0.59148$. The resulting curve for 
the given set of parameters is plotted in \figref{fig:BHcurve}. The plate conductivity has 
been set equal to the tabulated experimental value for the given steel grade, namely 
$\cond=6.993~MS/m$. The strip thickness is $d=2$~mm, which corresponds to a typical thickness 
of steel strips produced for the auto-mobile industry.

The two coils are identical with inner and outer radius $\rho_i = 10$~mm and $\rho_o = 20$~mm,
respectively, length $l=10$~mm, and are wound with 336 turns each. The lift-off is taken equal
to $0.5$~mm, i.e. the coils are considered being in contact with the specimen. The working 
frequency is $60$~Hz.

Given the fact that the modal solution is by construction valid only during the steady-state 
regime, it is expected that the two results will differ during the transient response of the 
system. Therefore, the FIT solution is calculated for two periods, and the results are compared 
only during the second period, where the system has reached the steady-state.

\begin{figure}[h]
\centering
\includegraphics[width=\plotwidth]{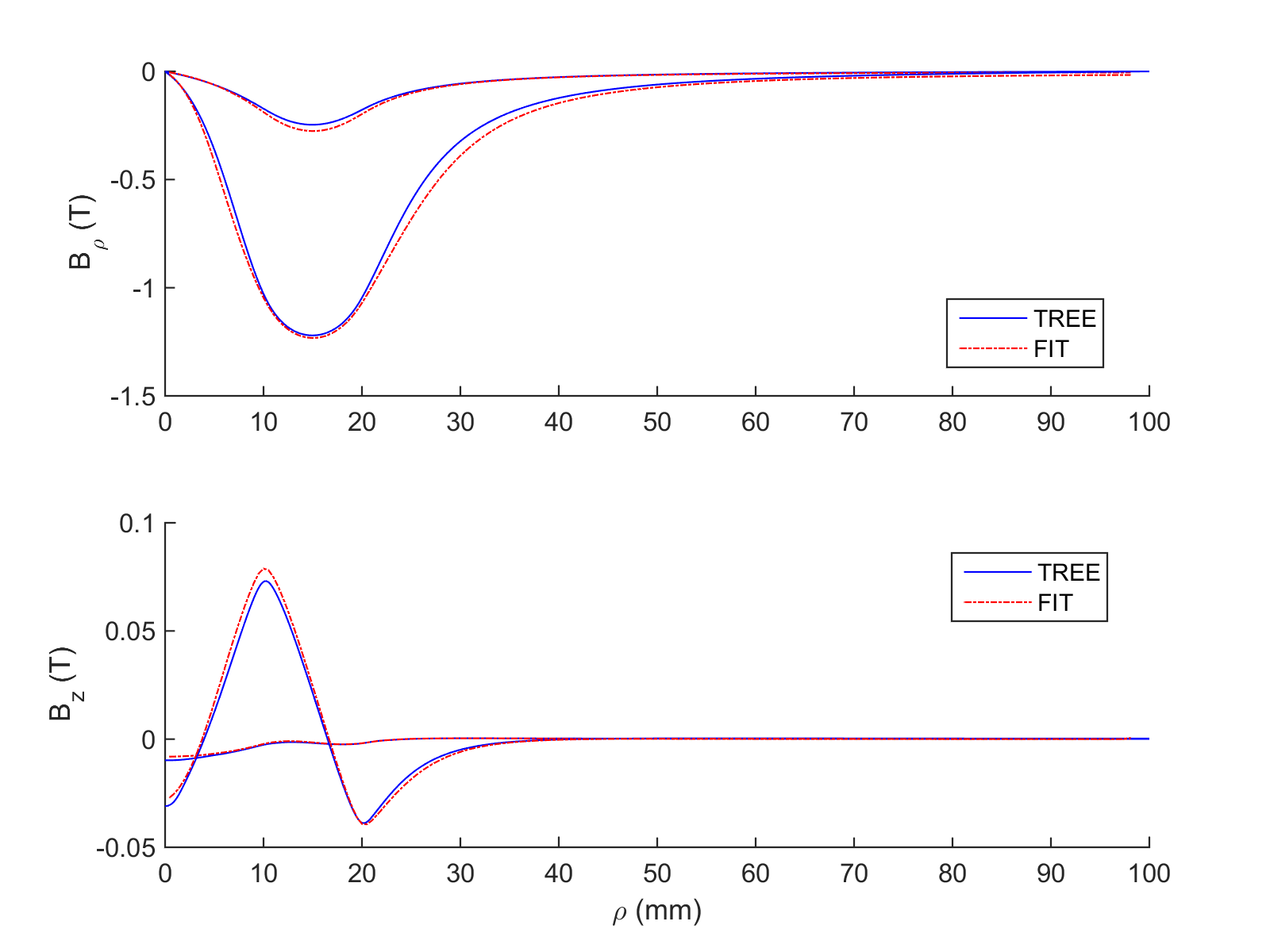}
\caption{Modal (TREE) vs. numerical (FIT) solution inside the plate as a function of the 
radial distance and at constant depth (results for 3 and 10~A). The ploted lines correspond 
to the field values at a given time instance.}
\label{fig:B_vs_r_10A}
\end{figure}

\figref{fig:B_vs_r_10A} shows the comparison between modal (TREE) and numerical (FIT) solution 
for the $B_\rho$ and $B_z$ variation with the distance from the coils axis at a depth equal to 
$d/4$ and $t=2$~ms. The two sets of curves correspond to excitation currents of 3 and 10~A.

\begin{figure}[h]
\centering
\includegraphics[width=\plotwidth]{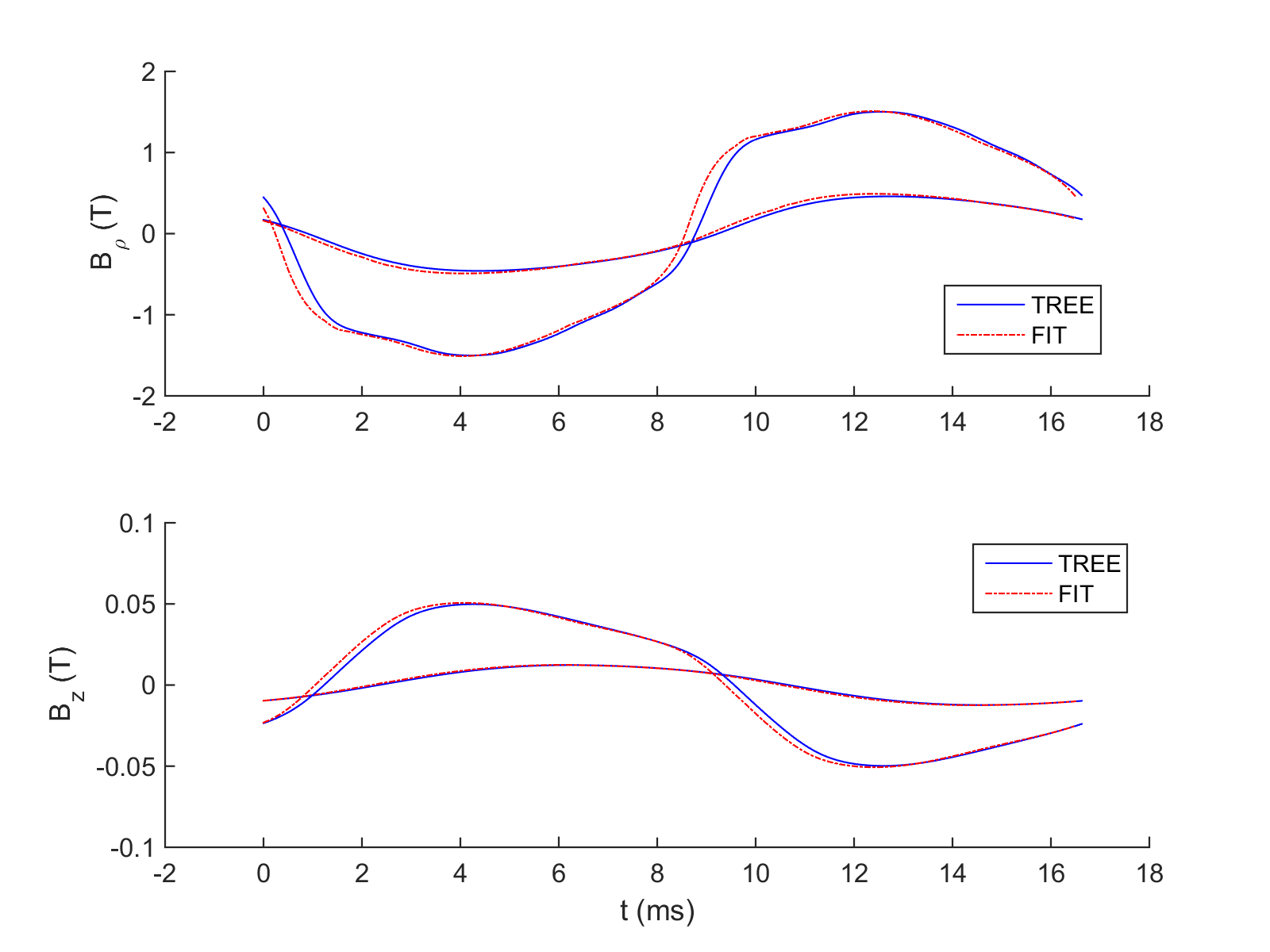}
\caption{Modal (TREE) vs. numerical (FIT) solution under the coil as a function of time
at the point $((\rho_i+\rho_o)/2,\;d/4)$. The two sets of curves correspond to the results
obtained for 3 and 10~A.}
\label{fig:B_vs_t_10A}
\end{figure}

The comparison for the temporal variation of both field components at radial distance of
$\rho = (\rho_i+\rho_o)/2$ and at the same depth with before ($d/4$) is shown in 
\figref{fig:B_vs_t_10A}. The specific observation point has been selected since it is the
location where the induced field reaches its maximum amplitude, and hence the non-linear 
effect becomes more profound (cf. \figref{fig:B_vs_r_10A}).

\begin{figure}[h]
\centering
\includegraphics[width=\plotwidth]{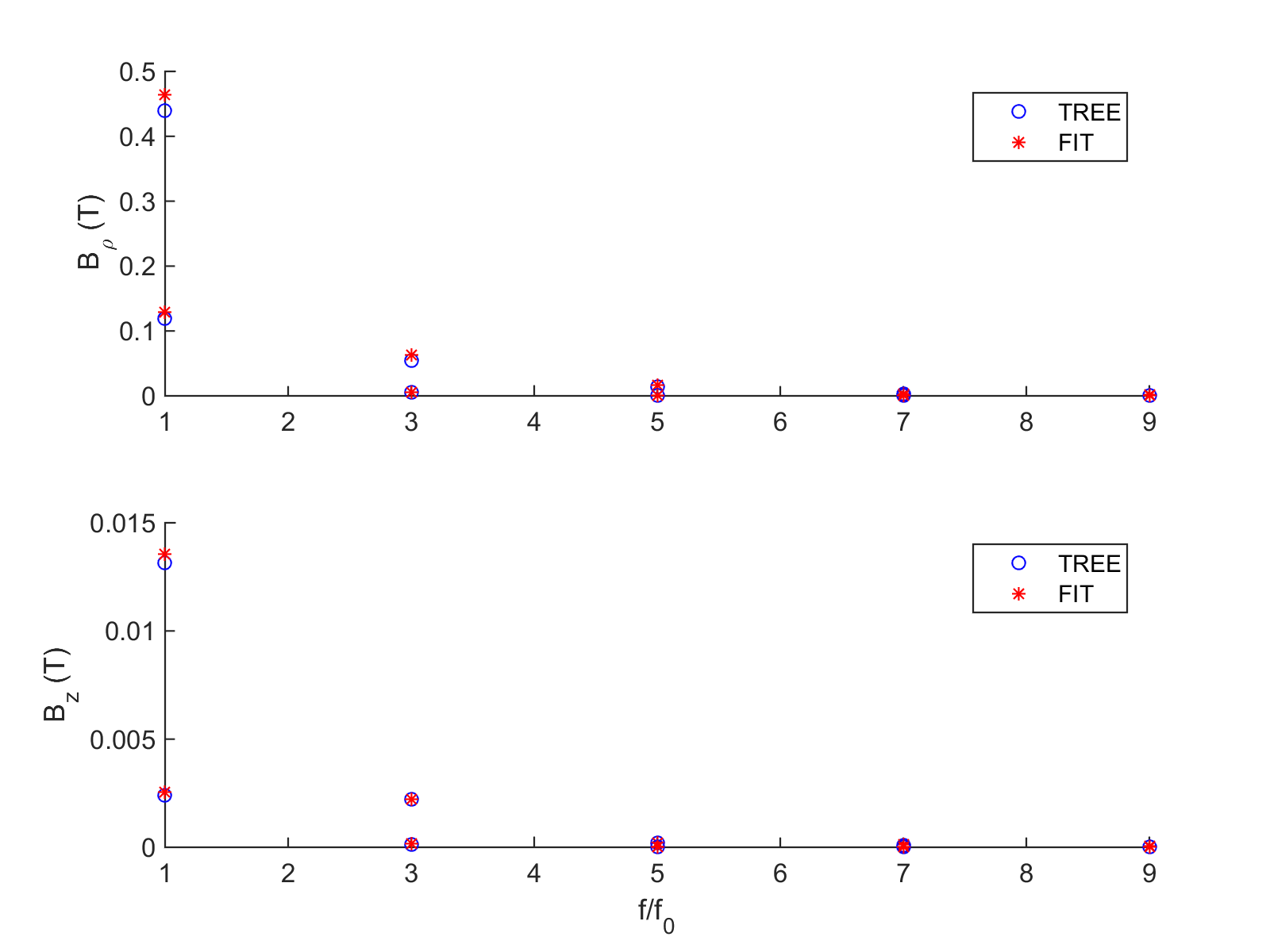}
\caption{Comparison of the spectra for the modal (TREE) and the numerical (FIT) solution 
at the point $((\rho_i+\rho_o)/2,\;d/4)$. Results obtained for 3 and 10~A.}
\label{fig:B_spec_10A}
\end{figure}

It is interesting to illustrate the harmonic content of the solution for the selected 
observation point and calculate the harmonic distortion since it offers a measure of the 
deviation from the linear regime. It also visualises the contribution from the different 
harmonics giving feedback where the spectrum can be truncated without loss of information. 
Finally, the harmonic distortion presents a practical interest since it is one of the more 
common measurements made by industrial devices of material characterization applications. 
The comparison of the spectra for the results of the two excitations is shown in 
\figref{fig:B_spec_10A}.

A common definition for the harmonic distortion reads
%
\begin{equation}
K = \sqrt{\frac{\sum_{p=1}^{\infty} A_{2p+1}^2}{A_1^2}}.
\label{eq:HrmDistDef}
\end{equation}
where $|A_p|$ is the amplitude of the $p$th harmonic. The comparison of the values for the
harmonic distortion factor calculated using the modal and the numerical solution for the
two components of the magnetic induction at the considered observation point is given in
\tabref{tab:HrmDistComp}.

\begin{table}[h]
\centering
\caption{Comparison of the harmonic distortion factor calculated using the modal and the 
numerical solution}
\begin{tabular}{c|c|c|c|c}
& \multicolumn{2}{c|}{$I=$3A} & \multicolumn{2}{c}{$I=$10A} \\
\hline
& $B_r$ & $B_z$ & $B_r$ & $B_z$ \\
\hline
TREE & 0.042 & 0.052 & 0.127 & 1.169 \\
FIT & 0.050 & 0.070 & 0.141 & 1.166
\end{tabular}
\label{tab:HrmDistComp}
\end{table}

\section{Conclusions}

The modal approach can be successfully applied to address the non-linear induction problem in
ferromagnetic specimens with canonical geometry. If, in addition, the considered piece is 
infinitely long, e.g. in the case of an infinite multilayer planar or cylindrical specimen,
the linearised operator is diagonalisable, which makes the non-linear iterations very cheap.

The presented analysis was restricted to harmonic excitations, and the fact that only a small
number of higher harmonics are excited has been exploited for reducing the computational cost.
The extension to excitations with periodic (power) signals is straight-forward. The more
complex problem of the transient response calculation as well as the response to finite 
duration (energy) signals, although amenable to similar treatment using Fourier transform, can
be more efficiently treated in time or Laplace domain. The corresponding analysis is under way
and will be presented in a separate article.

\section{Acknowledgement}

This work is supported by the CIVAMONT project, aiming at developing scientific collaborations
around the NDT simulation platform CIVA developed at CEA LIST.


\bibliographystyle{elsarticle-num}

\end{document}